\documentclass{article}

\usepackage[english]{babel}
\usepackage[a4paper]{geometry}

\usepackage{amsmath}
\usepackage{graphicx}
\usepackage{authblk}
\usepackage[colorlinks=true, allcolors=blue]{hyperref}

\title{ScopeOne: Flexible and C++-driven
Microscope Control Platform}
\author[1]{Tianyi Zhao}
\author[1,2]{Staffan Persson}
\author[1,3,*]{Guillermo Moreno-Pescador}
\affil[1]{Department of Plant and Environmental Sciences, University of Copenhagen, 1871 Frederiksberg C, Denmark}
\affil[2]{Copenhagen Plant Science Centre, University of Copenhagen, 1871 Frederiksberg C, Denmark}
\affil[3]{Niels Bohr Institute, University of Copenhagen, 2200 Copenhagen N, Denmark}
\affil[*]{Corresponding author: moreno@plen.ku.dk}

\begin{document}
\maketitle

\begin{abstract}
Modern microscopy systems integrate heterogeneous hardware devices that require dedicated software for coordination. However, high-performance C++ implementations of microscopy control software remain scarce. We present ScopeOne, a C++ and Qt-based microscopy control software built on the $\mu$Manager hardware abstraction layer. Through process isolation and shared memory, ScopeOne achieves simultaneous multi-camera preview with real-time image processing, while maintaining full compatibility with the $\mu$Manager device ecosystem.
\end{abstract}

\section{Introduction}
Custom microscopes are typically assembled from cameras, stages, illumination hardware, shutters, and other specialized components sourced from different vendors. Managing this heterogeneous hardware has traditionally meant relying on vendor-specific applications, which are often closed-source and difficult to extend across mixed hardware environments. Alternatively, commercial platforms such as LabVIEW and MATLAB offer broader integration capabilities but require costly licenses. The open-source microscopy community has made remarkable progress in addressing hardware fragmentation. $\mu$Manager\cite{edelstein_computer_2010} has established a mature ecosystem with device adapters for various types of hardware. Pycro-Manager\cite{pinkard_pycro-manager_2021}, pymmcore\cite{noauthor_micro-managerpymmcore_2026} and pymmcore-plus\cite{noauthor_pymmcore-pluspymmcore-plus_2026} extend this foundation with Python interfaces that enable integration into the rich ecosystem of computer vision and image processing libraries. Tormenta \cite{barabas_note_2016} and ImSwitch \cite{moreno_imswitch_2021} illustrate the shift toward modular, Python-based control systems that can support multiple microscopy modalities within a unified framework.

While existing open-source microscopy software provides useful building blocks, ScopeOne is designed for a specialized and relatively unaddressed area of multi-camera real-time preview and live image processing. Cameras produce continuous high-bandwidth data streams, and on-the-fly processing such as background subtraction and filtering is essential for alignment and optimization. In these scenarios, C++ provides direct memory control and minimal runtime overhead for handling high-throughput data pipelines.  A common limitation among open-source microscope control software is fragmented device support, as software packages typically maintain their own device adapters, resulting in duplicated work across the community. ScopeOne builds on MMCore's hardware abstraction layer, which allows it to support an extensive range of hardware while keeping development focused on the performance-critical preview and processing components.

\section{Implementation}
\begin{figure}[h]
        \centering
        \includegraphics[width=1\linewidth]{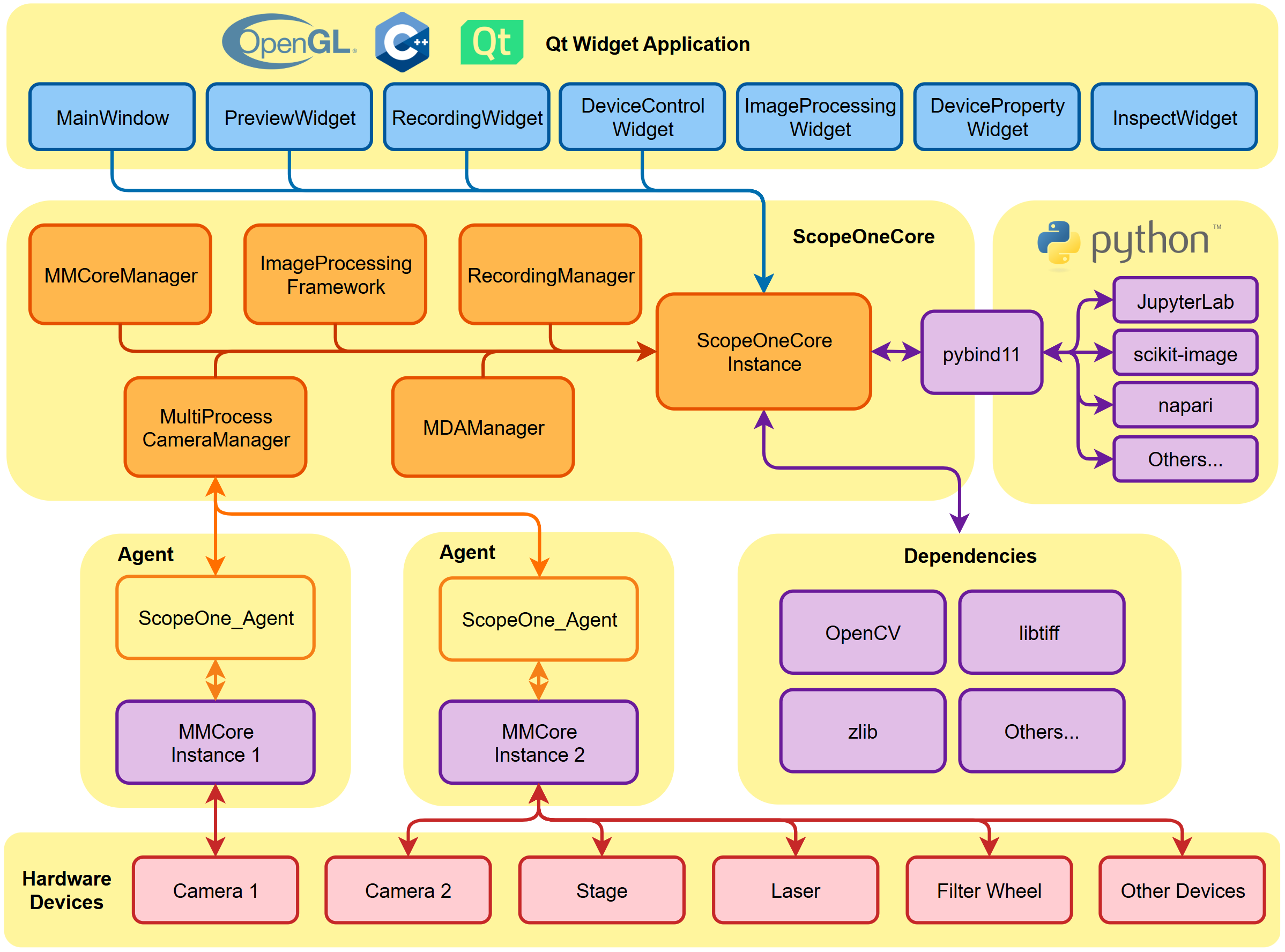}
        \caption{ScopeOne architecture}
        \label{a}
\end{figure}
ScopeOne is built on Micro-Manager's MMCore, which abstracts heterogeneous microscope hardware through a unified control interface. On this foundation, ScopeOne adopts a three-layer architecture consisting of a hardware abstraction layer based on MMCore, a core service layer implemented as the shared library ScopeOneCore, and an interface layer. ScopeOneCore encapsulates the main runtime logic, including device coordination, camera acquisition, image processing, and recording, and exposes a unified API used by both a Qt Widgets desktop application and Python bindings implemented via pybind11. This allows the desktop application and Python bindings to share the same underlying logic.

ScopeOne uses MultiProcessCameraManager as a unified acquisition manager for both single-camera and multi-camera operation. In single-camera configurations, it acquires images directly through the host MMCore instance. In multi-camera configurations, MMCore can work with multiple cameras simultaneously only if their image sizes are identical\cite{chhetri_software_2020}. Therefore, ScopeOne launches an independent ScopeOneAgent for each camera, each agent maintaining its own MMCore instance and camera state. Frames from each agent are delivered to the host through shared memory, providing process-level isolation while keeping acquisition centrally orchestrated.

The ImageProcessingFramework enables real-time processing of live camera streams without interrupting image acquisition. It adopts a modular pipeline architecture in which processing modules are configured sequentially and can be added or removed as needed. The current implementation supports FFT bandpass filter, median filter, Gaussian filter, background calibration, and spatiotemporal binning.  The framework leverages mature OpenCV algorithms for image processing, while the processing pipeline is executed asynchronously using Qt’s multithreaded framework to maintain responsive preview and recording. The processed output is immediately displayed in the user interface, allowing for rapid optimization of imaging conditions during experiments.
\begin{figure}
    \centering
    \includegraphics[width=1\linewidth]{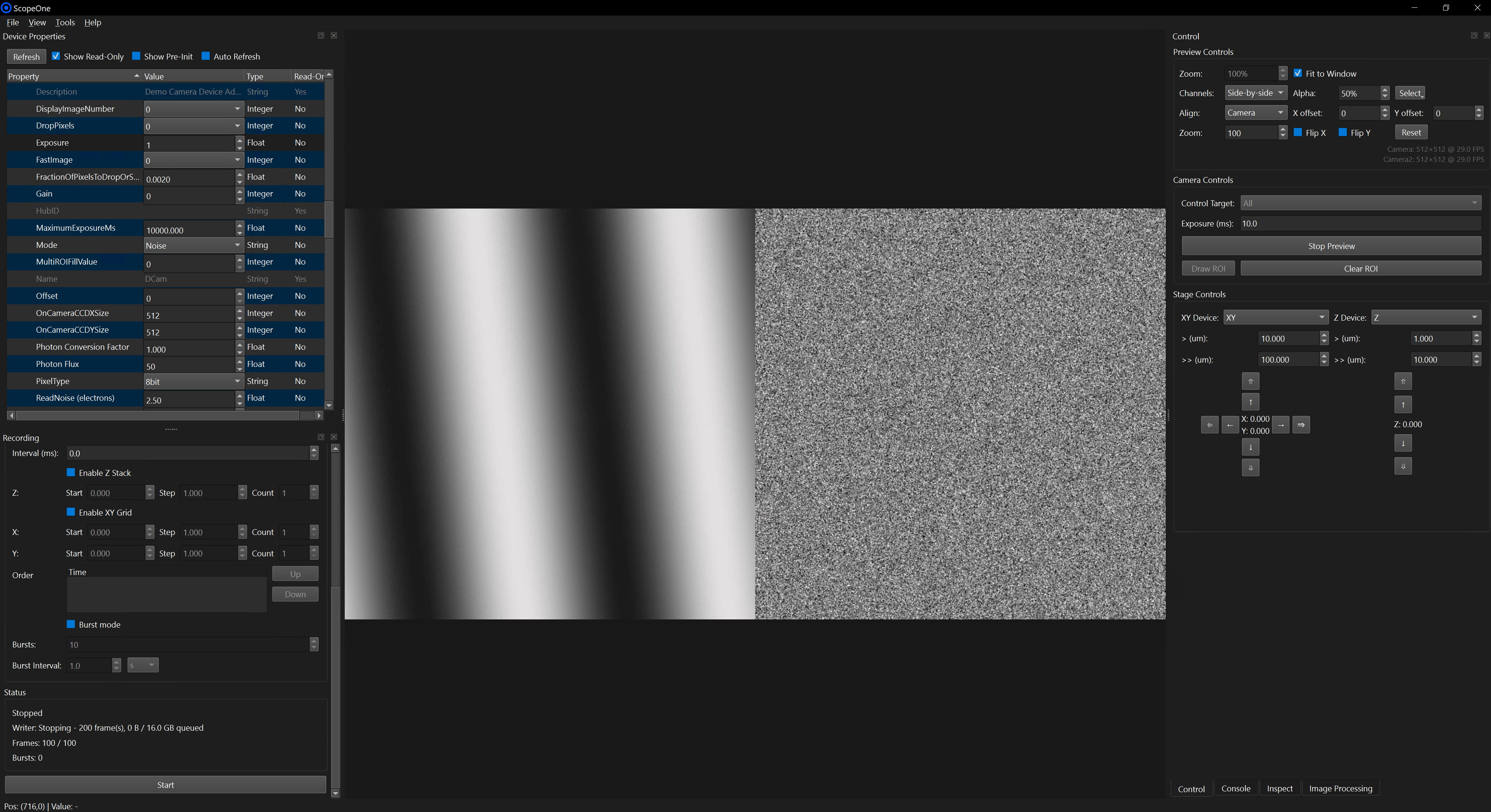}
    \caption{ScopeOne graphical user interface. Left side panel gathers hardware and recording settings. Central panel shows the cameras views and processing outputs. Right panel shows the GUI controls for: Camera, preview and stage.}
    \label{u}
\end{figure}

ScopeOne employs an OpenGL-based preview widget to provide a smooth and responsive live preview. Incoming raw and processed frames are uploaded to the GPU as textures and rendered through a lightweight shader pipeline. The preview supports both side-by-side and overlay display modes, allowing multiple camera streams or raw/processed outputs to be compared within a unified interface. It also provides interactive operations such as zooming, per-channel display adjustment, ROI selection, histogram equalization, and cross-section live visualization, enabling responsive visual inspection during acquisition.

ScopeOneCore exposes its API to Python through pybind11, enabling programmatic control and automation of microscopy workflows. The Python module provides direct access to image acquisition, processing, and multi-dimensional recording. This allows users to write scripts for automated experiments, batch processing, or integration with data analysis pipelines. It also integrates with widely used scientific libraries such as NumPy, SciPy, napari, and Jupyter notebooks, supporting analysis and reproducible research workflows.

\section{Conclusion}

ScopeOne presents a high-performance, open-source microscopy control solution that addresses the limitations in existing tools in multi-camera scenarios. Built on a C++ and Qt-based architecture centered on the extensible ScopeOneCore library, it supports simultaneous preview from multiple cameras, real-time image processing, and compatibility with the Micro-Manager device ecosystem. Python bindings allow both GUI operation and programmatic automation, making it suitable for a range of experimental workflows.

As an open-source project, ScopeOne builds on existing community efforts to reduce duplicated work and provides an alternative that enriches and diversifies the microscopy community. Although current development is conducted in close collaboration with the optics and biology teams within our laboratory, we aim to expand engagement with the broader research community to make the platform more practical, accessible, and universal. Contributions in the form of \href{https://github.com/Experimental-Microscopy-Lab/ScopeOne/issues}{issues} and \href{https://github.com/Experimental-Microscopy-Lab/ScopeOne/pulls}{pull requests} are greatly appreciated.

\section{Acknowledgment}
We thank all developers and contributors to the open-source microscopy software community for the resources and inspiration that supported the early development of ScopeOne. In particular, Micro-Manager provided the foundation for ScopeOne and helped us understand how the graphical user interface interacts with the MMCore hardware abstraction layer. We also acknowledge the ImSwitch and Tormenta projects, which inspired aspects of our graphical user interface layout design.

The authors also wish to acknowledge the support of the Novo Nordisk Foundation funding (grants numbers NNF22OC0079093 and NNF24OC0088090).

\section{Data availability}
The source code can be found at \href{https://github.com/Experimental-Microscopy-Lab/ScopeOne}{https://github.com/Experimental-Microscopy-Lab/ScopeOne}.

\bibliographystyle{unsrt}
\bibliography{ref}

\end{document}